\documentclass[a4paper,conference]{IEEEtran}
\ifCLASSINFOpdf
\else
\fi
\IEEEoverridecommandlockouts
\usepackage{cite}
\usepackage{amsmath,amssymb,amsfonts} 
\usepackage{algorithmic}
\usepackage{graphicx}
\usepackage{textcomp}
\usepackage{xcolor}
\usepackage{booktabs}
\usepackage{subfigure} 
\usepackage{mathrsfs}
\usepackage{multirow}
\def\BibTeX{{\rm B\kern-.05em{\sc i\kern-.025em b}\kern-.08em
    T\kern-.1667em\lower.7ex\hbox{E}\kern-.125emX}}

\begin{document}

\title{Multi-Graph Convolutional Network for Relationship-Driven Stock Movement Prediction}
\author{
\IEEEauthorblockN{Jiexia Ye, Juanjuan Zhao*, Kejiang Ye}
\IEEEauthorblockA{
\textit{Shenzhen Institutes of Advanced Technology} \\
\textit{Chinese Academy of Sciences}\\
Shenzhen, China \\
\textit{University of Chinese Academy of Sciences}\\
Beijing, China \\
Email: \{jx.ye,jj.zhao,kj.ye\}@siat.ac.cn
}
\and
\IEEEauthorblockN{Chengzhong Xu, IEEE Fellow}
\IEEEauthorblockA{
\textit{State Key Lab of IOTSC, Department of Computer Science} \\
\textit{University of Macau}\\
Macau SAR, China \\
Email: \{czxu\}@um.edu.mo
}

\thanks{*Corresponding author: Juanjuan Zhao}
}

\maketitle
\begin{abstract}
Stock price movement prediction is commonly accepted as a very challenging task due to the volatile nature of financial markets. Previous works typically predict the stock price mainly based on its own information, neglecting the cross effect among involved stocks. However, it is well known that an individual stock price is correlated with prices of other stocks in complex ways. To take the cross effect into consideration, we propose a deep learning framework, called Multi-GCGRU, which comprises graph convolutional network (GCN) and gated recurrent unit (GRU) to predict stock movement. Specifically, we first encode multiple relationships among stocks into graphs based on financial domain knowledge and utilize GCN to extract the cross effect based on these pre-defined graphs. To further get rid of prior knowledge, we explore an adaptive relationship learned by data automatically. The cross-correlation features produced by GCN are concatenated with historical records and then fed into GRU to model the temporal dependency of stock prices. Experiments on two stock indexes in China market show that our model outperforms other baselines. Note that our model is rather feasible to incorporate more effective stock relationships containing expert knowledge, as well as learn data-driven relationship.

\end{abstract}

\IEEEpeerreviewmaketitle
\section{Introduction}
\label{introduction}
Predicting the future status of a stock has always been of great interest by many investors for that a little improvement of prediction accuracy might yield a huge gain. Both traditional finance and modern behavior finance believe that fluctuations of stock prices are information-driven. Information affects beliefs and behaviors of investors, thus changing stock movement. Therefore, understanding how stock market impounds information into stock prices is paramount in stock prediction\cite{HouIndustry}. 

Recently, researches make substantial effort on modeling correlations between various information and stock prices by machine learning\cite{DBLP:journals/asc/LuoYXP17},\cite{DBLP:journals/asc/ChangL17},\cite{jin2017tracking},\cite{DBLP:conf/ictai/ChangT18} or deep learning\cite{feng2019enhancing},\cite{DBLP:conf/ictai/LiuS17},\cite{zhang2017stock},\cite{chen2019investment}.  However, the core assumption of these algorithms in most works is that stocks are independent of each other. They mainly focus on extracting the autocorrelation of an individual stock based on its own historical information but neglect the cross effect of stocks over time, which affects the stock prices dynamically\cite{Andrew1990When}.

The cross interaction among stocks is attributed to various connections among corporations, such as the shared industry information\cite{HouIndustry}, supply chain, payment network\cite{DBLP:journals/epjds/LetiziaL19}, business partnership and shareholder ownership\cite{chen2018incorporating}. These complex and multifaceted connections result in that the price of an individual stock is highly correlated with other stocks besides its own information. For instance, considerable empirical findings have confirmed a pronounced lead-lag pattern in stock prices, i.e. some stock prices lead or lag other stock prices\cite{Andrew1990When},\cite{HouIndustry}. Therefore, it is natural to take the corporation relationships into consideration for better stock prediction. 

However, there are three major challenges in utilizing these relationships: (1)  design appropriate representation  for corporation relationships, (2) design a model without independent instance assumption to extract the cross-correlation among stocks, (3) predict the target stock movement by jointly considering its historical observation and the cross-correlation with related stocks.

To address the first challenge, we follow previous works \cite{DBLP:journals/epjds/LetiziaL19}, \cite{chen2018incorporating} to embed the corporation relationships into graphs. In each graph, a node represents a listed company and the edge represents interaction between two listed companies. Besides inheriting the shareholding graph in \cite{chen2018incorporating}, we novelly define an industry graph based on lead-lag theory \cite{Andrew1990When} and a topicality graph based on common topical news\cite{DBLP:conf/wsdm/HuLBLL18}. In addition, we realize that these artificial relational graphs depend on solid financial knowledge and require more financial data. To overcome such limitations, we establish a dynamic graph on the basis of data. We analyze the effectiveness of these graphs and make necessary comparison.

After establishing relational graphs among stocks, we propose a deep learning framework called Multi-GCGRU by jointly combining Graph Convolutional Network (GCN) and Gated Recurrent Unit (GRU) to tackle the second and the third challenges. GCN is a state-of-the-art deep learning approach to handle the complexity of graph data. It has demonstrated its effectiveness in capturing interdependency between instances in a graph and has achieved state-of-the-art performance in many applications, such as molecular fingerprints\cite{duvenaud2015convolutional}, recommendation system\cite{ying2018graph} and traffic forecasting\cite{cuitraffic19}. In this paper, we perform graph convolutions on the pre-defined graph structures to model various interactions among stocks. Since stock prediction is a time series task and GRU has been proved to be effective for processing sequential data\cite{chung2014empirical}, we utilize GRU to learn the temporal dependency from historical market data along with the cross-effect features produced by GCN. We test our Multi-GCGRU model on two real stock market indexes. The experimental results show that our model has better performance than baseline methods.

Specifically, our contributions can be summarized as follows:

(1) We take the cross effect among a collection of stocks into consideration for better stock prediction. We construct multiple graphs based on various corporation relationships to enrich the representations of cross effect. Further, to get rid of prior knowledge on financial domain, we explore an adaptive graph.

(2) We employ Graph Convolutional Network (GCN) to process these graphs along with historical information to learn complicated interactions among related stocks, and produce new features for each individual stock in the collection which contains cross-impact information from other stocks.

(3) We concatenate the cross-effect features with historical market information at the same time slice for each stock. These combined features are fed into Gated Recurrent Unit (GRU) to learn temporal pattern in stock prices.

(4) Note that our model can be easily extended to incorporate more effective relationships among stocks. Even without any pre-defined relationship, our model is able to learn a dynamic graph automatically from market price data. The code of Multi-GCCRU is publicly available from https://github.com/start2020/Multi-GCCRU.

\section{Related Works}
\subsection{Stock Price Prediction}
Stock price prediction is a very challenge task due to the diverse and complicate factors, including corporate financial performance\cite{DBLP:conf/sdm/KampBG14}, industry information\cite{HouIndustry}, public news\cite{CHAN2003223},\cite{DBLP:conf/wsdm/HuLBLL18}, social sentiment\cite{DBLP:conf/acl/RekabsazLBDAH17},\cite{DBLP:conf/ictai/SousaSRMFM19}. Recently, various traditional machine learning and deep learning approaches have been proposed to extract valuable clues from different types of information sources for better stock prediction.

The input features of most works are mainly based on historical market data (e.g. stock prices, trading volume). For instance, Zhang et al. \cite{zhang2017stock} utilized only historical prices to capture the multi-frequency trading patterns by a novel State Frequency Memory (SFM) recurrent network. Feng et al. \cite{feng2019enhancing} focused on addressing the stochasticity of stock price variable to improve the generalization of prediction model by proposing an adversarial training solution. However, only historical price data cannot entirely explain the volatility of stock price. Other types of information are complementary to enrich the input features and help to discover more concealed rules in stock price, such as public news\cite{DBLP:conf/wsdm/HuLBLL18}, \cite{CHAN2003223}, texts from social medias\cite{li2015tensor},\cite{xu2018stock} and web browsing data\cite{bordino2014stock}. For example, Ding et al. \cite{ding2015deep} extracted events from news titles to model influence of events on stock price by a CNN-based framework. Xu et al. \cite{xu2018stock} presented a novel deep generative model to learn opinions from Twitter texts. Cheng et al. \cite{chen2019investment} explored the mutual fund portfolio data to extract stock intrinsic properties for enhancing prediction. Qin \cite{qin2019you} integrated CEO's vocal features in a conference call into the model.

Despite tremendous efforts have been made to understand the principle of stock price movement,  most of the works above mainly focused on combining a single stock's historical records with other textual information but overlooked the correlations among stocks. Only a few attempts \cite{li2019multi} have been made to explore the cross effect among stocks which has been verified by\cite{Andrew1990When}. In this paper, we pay attention to model the influences from other stocks on the target stock.

\subsection{Graph Convolutional Network}
In the last couple of years, many attempts have been made to generalize neural networks on graph-structured data. Encouraged by the success of CNN, researchers have successfully re-defined the notion of convolution on graph data, called graph convolution. The corresponding neural network, i.e. graph convolution network (GCN) has gained much attention recently since it has demonstrated outstanding performance on node classification task\cite{kipf2017semi}. GCN takes the graph structure and node features as inputs and it can capture the complex interaction between nodes on graph by aggregating information from neighbors and doing non-linear transformation on all feature dimensions to create new features. Such capacity enables GCN to achieve state-of-the-art performance in graph related applications \cite{9207049},\cite{ying2018graph},\cite{DBLP:journals/corr/abs-2005-11691}. In recent research, Chen et al. \cite{chen2018incorporating} applied GCN in stock prediction and they modeled the correlations among stocks based on a shareholding graph. However, such graph is rather limited due to the sparse cross-shareholdings among public corporations and it is insufficient to represent the complex correlations among stocks. We would compare our model with this method in experiments.

\section{Problem Formulation}
Stock price prediction can be divided into stock price return prediction\cite{chen2019investment} and stock movement prediction\cite{li2019multi},\cite{feng2019enhancing}, \cite{li2015tensor}, where the former predicts the exact price of stock while the latter predicts the up or down of stock price. Due to the complexity and stochasticity of stock market, it is rather difficult to predict price return but stock movement prediction is more achievable. Therefore, most works focus on stock movement prediction and so do we.

Usually, stock price movement aims to predict the movement of a target stock in a pre-selected stock collection on a target trading day with historical price information along with other information\cite{xu2018stock}. The mathematical formulation is as follows:

\begin{footnotesize}\begin{equation}
\hat{y}^{\mathbf{s}}_{\mathbf{d}}=f([x^{\mathbf{s}}_{\mathbf{d}-\mathbf{P}},\cdots, x^{\mathbf{s}}_{\mathbf{d}-1}],\mathbf{E};\Theta)
\end{equation}\end{footnotesize}where $\mathbf{s}$ is the target stock,  $\mathbf{d}$ is the target day, $\mathbf{P}$ is the lag size, $x^{\mathbf{s}}_{t} \in \mathbb{R}^{\mathbf{F}}$ is the $\mathbf{F}$ historical features of the target stock at day $t$, $\mathbf{E}$ is external information, $\Theta$ is trainable parameter. $\hat{y}^{\mathbf{s}}_{\mathbf{d}}\in [0,1]$ is the predicted probability at day $\mathbf{d}$.

\begin{figure}[htb]
\centering
\includegraphics[width=0.45\textwidth]{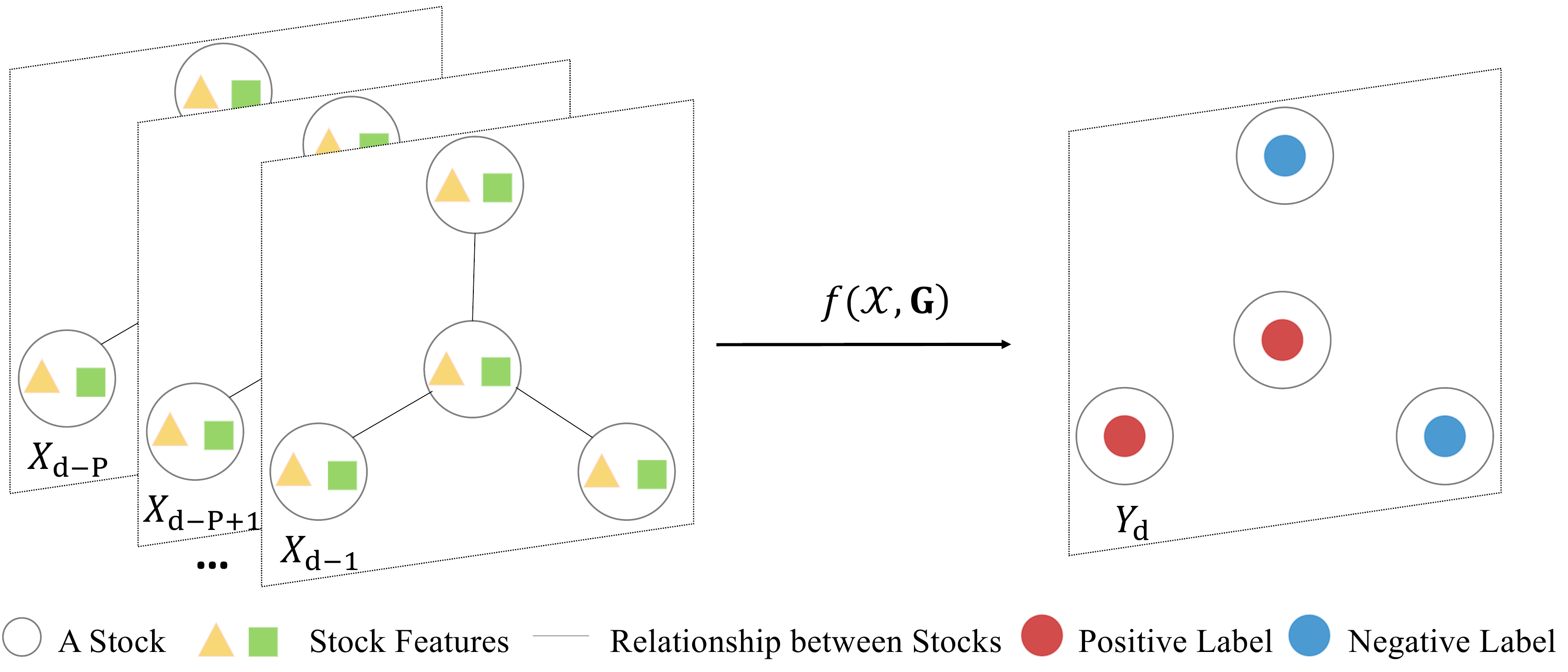}
\caption{Problem Formulation of Stock Movement Prediction Based on Graph}
\label{fig:Formulation}
\end{figure}

However, such formulation treats each stock independently and overlooks its complex correlations with other related stocks. We encode various correlations among stocks as graphs and focus on exploring relationship-driven influence for stock prediction (as shown in Figure \ref{fig:Formulation}). Therefore, we re-formalize the problem as follows:

\begin{footnotesize}\begin{equation}
\hat{Y}_{\mathbf{d}}=f([X_{\mathbf{d}-\mathbf{P}},\cdots, X_{\mathbf{d}-1}],\mathbf{G};\Theta)
\end{equation}\end{footnotesize}where $X_{t}\in \mathbb{R}^{\mathbf{N} \times \mathbf{F}}$ denotes a snapshot of stock collection $\mathbf{S}$ with $\mathbf{N}$ stocks at day $t$. $\mathbf{G}$ is the graph. $\hat{Y}_{\mathbf{d}}=[\hat{y}^{1}_{\mathbf{d}},\cdots,\hat{y}^{\mathbf{N}}_{\mathbf{d}}] \in \mathbb{R}^{\mathbf{N}}$ is the predicted series labels of collection $\mathbf{S}$ at day $\mathbf{d}$. We use cross entropy function as the loss function:

\begin{footnotesize}\begin{equation}
\mathcal{L}=-\frac{1}{\mathbf{N}} \sum_{\mathbf{s}=1}^{\mathbf{N}}[y_{\mathbf{d}}^{\mathbf{s}}log(\hat{y}_{\mathbf{d}}^{\mathbf{s}})+(1-y_{\mathbf{d}}^{\mathbf{s}})log(1-\hat{y}_{\mathbf{d}}^{\mathbf{s}})]\end{equation}\end{footnotesize}where $Y_{\mathbf{d}}=[y^{1}_{\mathbf{d}},\cdots,y^{\mathbf{N}}_{\mathbf{d}}]$ is denoted as the ground truth series and $y^{\mathbf{s}}_{\mathbf{d}}\in \{0,1\}$ for that most works estimate the binary movement with $1$ denoting as rise or positive, $0$ denoting as fall or negative \cite{feng2019enhancing}, \cite{li2015tensor}.

\section{Multi-GCGRU}
\begin{figure*}[htb]
\centering
\includegraphics[width=1.0\textwidth]{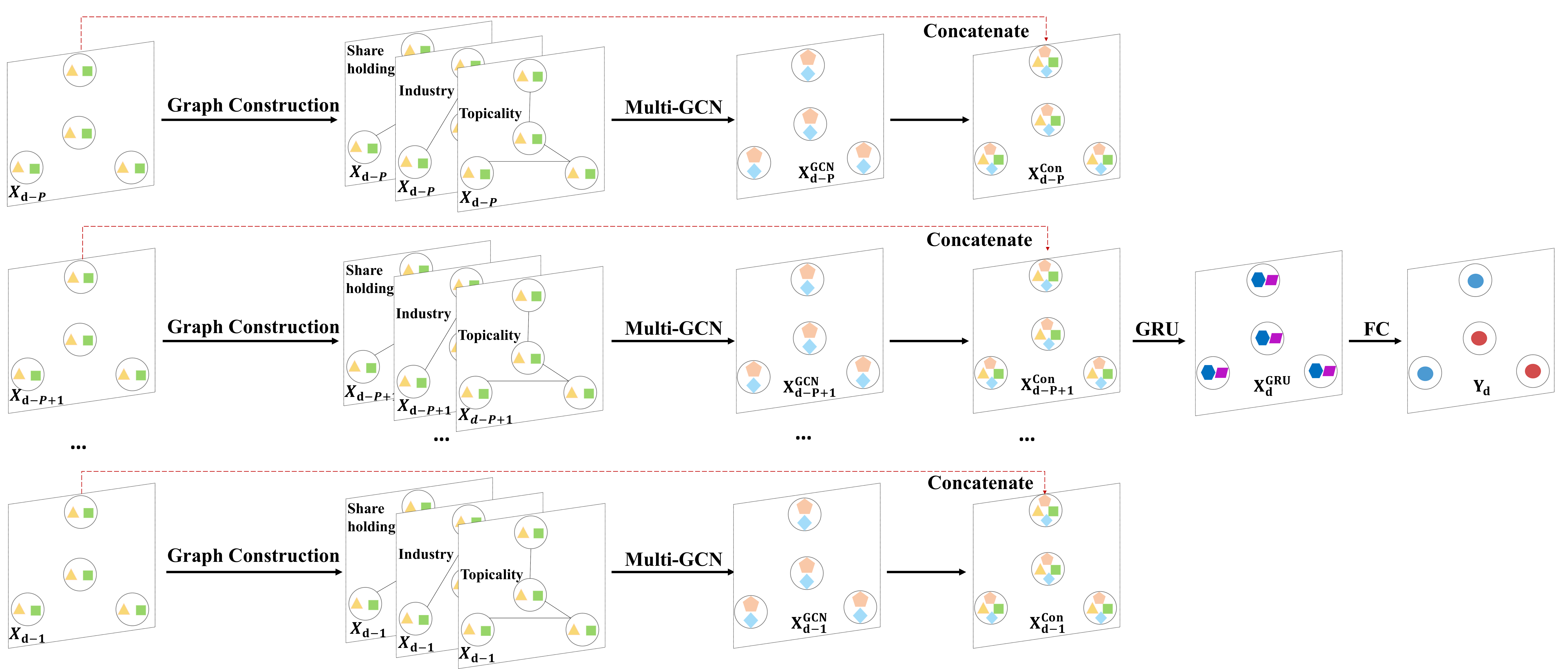}
\hspace{-2.58mm}
\vspace{1em}
\caption{The architecture of Multi-GCGRU; GCN: Graph Convolution Network; GRU: Gated Recurrent Unit; FC: Fully connected layer. $X^{\mathbf{GCN}}_{t}$ is the output of Multi-GCN at day $t$. $X^{\mathbf{Con}}_{t}$ is the concatenated features based on $X^{\mathbf{GCN}}_{t}$ and the historical records of stock collection $X_{t}$ at time $t$. $X^{\mathbf{GRU}}_{\mathbf{d}}$ is the final output of GRU at target day $\mathbf{d}$.}
\label{fig:Model}
\end{figure*}

The Multi-GCGRU architecture in our paper aims to predict the stock price movement by considering both historical records of individual stock and cross effect from other related stocks. The first stage is to extract relationships which can explain the cross effect among stocks by encoding them into graphs (e.g. Shareholding Graph, Industry Graph, Topicality Graph). The second stage is to learn more complicated and dynamic cross-correlation behind stock collection by Multi-GCN. The third stage is to utilize GRU to learn the temporal dependency by historical records along with higher features containing cross effect produced by Multi-GCN. Finally, a fully connected layer with sigmoid activation function is added to get the probability prediction.  The details of the architecture are shown in Figure \ref{fig:Model}.

\subsection{Graph Construction}

The cross-autocorrelation of stock returns over time has long been recognized in stock markets\cite{Andrew1990When}. To capture the complex cross effect among a clique of stocks, we extract three types of relationships based on prior financial knowledge and construct three graphs respectively, including (1) Shareholding Graph $\mathbf{G_{S}} = (\mathbf{V},\mathbf{E_{S}},\mathbf{A_{S}})$ to encode shareholding influence\cite{chen2018incorporating}, (2) Industry Graph $\mathbf{G_{I}} = (\mathbf{V},\mathbf{E_{I}},\mathbf{A_{I}})$ to encode lead-lag effect within industry\cite{Andrew1990When}, (3) Topicality Graph $\mathbf{G_{T}} = (\mathbf{V},\mathbf{E_{T}},\mathbf{A_{T}})$ to encode topical news impact\cite{CHAN2003223}, where $|\mathbf{V}|= \mathbf{N}$ refers $\mathbf{N}$ public corporations in the graph, $\mathbf{A}=(\mathbf{a}_{ij})_{\mathbf{N}\times \mathbf{N}}$ is the adjacency matrix representing the particular stock network. Element $a_{ij}$ in $\mathbf{A}$ stands for the connection strength between company $i$ and company $j$. We examine the effectiveness of these relationships in our experiments.

\subsubsection{\textbf{Shareholding Graph}}
Chen et al. \cite{chen2018incorporating} defined a weighted shareholding graph based on the financial fact that the performance of a listed company is likely to influence the stock price of its shareholder which is also a listed company and vice versa\cite{ATTIG20062875}. The mutual influence strength depends on the shareholding ratio. Therefore, an edge is attached to two listed corporations which have shareholding relationship. The edge weight $a_{ij}$ is the shareholding ratio in range of $[0,1]$. However, we find that the cross-shareholdings among public corporations are rather rare empirically, which leads to a very sparse adjacent matrix and weakens its effective representation for the cross-effect among stocks.

\subsubsection{\textbf{Industry Graph}}
The cross effect among stocks has displayed a pronounced lead-lag structure\cite{Andrew1990When}, which refers that returns on some stocks systematically lead or lag returns on other stocks \cite{Hoffmann2013426}. The main cause of lead-lag effect in equity market can be explained by industry information diffusion hypothesis\cite{HouIndustry}, which argues that new information is usually incorporated into the stock prices of industry leaders before it spreads to other firms in the same industry. Therefore, lead-lag effect is related to the firm size and the stock returns of larger firms generally lead those of smaller ones\cite{Andrew1990When} in the same industry. 

In this paper, we construct an industry graph to model the lead-lag relationship. Since Hou et al. \cite{HouIndustry} has verified that the cross-industry lead-lag effect is rather weak, we mainly focus on intra-industry lead-lag effect. If two companies are in different industries, there is no edge between them. Otherwise, the influence from company $i$ to company $j$ is denoted as $a_{ij}=\frac{\mathbf{M_{i}}}{\mathbf{M_{j}}}$, where $\mathbf{M}$ denotes the firm size. Note that such influence is asymmetric. Returns of small firms are correlated with past returns of big firms, but not vice versa\cite{HouIndustry}. Since several studies \cite{Andrew1990When} have shown that stock with larger capital size tends to be a leading stock, we use registered capital to measure firm size in this paper.

\subsubsection{\textbf{Topicality Graph}}
The rapid development of Internet has accelerated the speed of producing and broadcasting news, enhancing the news impact on investment behaviors. Extensive studies have been conducted on the correlations between news and stock prices\cite{CHAN2003223},\cite{DBLP:conf/wsdm/HuLBLL18}. On one hand, it can be observed that a stock responses to a group of similar news with the same topicality. For example, the news related with 2019-nCoV is likely to impact a pharmaceutical stock. On the other hand, the news belonging to one particular topicality impact many related stocks, leading to the similar volatility of stock prices. For instance, news related with 2019-nCoV have influenced pharmaceutical stocks (e.g. ABIO), entertainment stocks (e.g. Disney) and caterers stocks (e.g.YUM).  

In this paper, we novelly define a topicality graph to model the correlation among stocks due to topical news impact. The stock datasets we collect from a public API (https://tushare.pro/) have already contained topical information for each stock, where a stock has more than one topicality and a topicality is shared by more than one stock. Based on these datasets, we measure the connection strength by the number of topicalities shared by two listed corporations. Intuitively, the more topicalities corporations share, the more similar their prices volatility might be. Specifically, if company $i$ owns $\mathbf{M_i}$ topicalities, company $j$ owns $\mathbf{M_j}$ topicalities and they share $\mathbf{T_{ij}}$ topicalities, the connection strength from $i$ to $j$ is denoted as $a_{ij}=\frac{\mathbf{T_{ij}}}{\mathbf{M_i}}$. Similarly, the connection strength from $j$ to $i$ is denoted as $a_{ji}=\frac{\mathbf{T_{ij}}}{\mathbf{M_j}}$. Note that if there is no topicality shared by two companies, their connection strength is zero.

\subsection{Multi-Graph Convolutional Network}
In this section, we introduce the architecture of Multi-Graph Convolutional Network (Multi-GCN) in our paper. We first introduce the traditional graph convolutional layer proposed by \cite{DBLP:conf/nips/DefferrardBV16}.  Then we novelly propose its variant, i.e. multi-graph convolutional layer, to incorporate multiple pre-defined stock graph structures into our model. As an alternative, we also design a dynamic graph convolutional layer to learn graph topology from data automatically in case that we don't have sufficient domain knowledge and data to pre-define a stock graph.  Finally,  the mathematical formalization of our Multi-GCN is defined.

\subsubsection{\textbf{Graph Convolutional Layer}}
When processing graph data, researchers hope to extract high level representation containing graph structure information. Inspired by convolution on images, Bruna et al. \cite{DBLP:journals/corr/BrunaZSL13} defined convolution on graph in spectral domain. The normalized graph Laplacian matrix $\mathbf{L} = \mathbf{I_{N}}-\mathbf{D}^{-\frac{1}{2}} \mathbf{A} \mathbf{D}^{-\frac{1}{2}} \in\mathbb{R}^{\mathbf{N} \times \mathbf{N}}$ is decomposed as $\mathbf{L}=\mathbf{U} \mathbf{\Lambda} \mathbf{U}^{T}$ to get a graph Fourier basis $\mathbf{U}$ which contains the graph topology information, where $\mathbf{D}$ is the degree matrix of $\mathbf{A}$, $\mathbf{U}$ is the eigenvectors matrix and $\mathbf{\Lambda}$ is the eigenvalues matrix. The graph signal $x\in \mathbb{R}^{\mathbf{N}}$ is first transformed to the spectral domain by the graph Fourier basis $\mathbf{U}$, then filtered by a parameterized kernel $\Theta$, finally transformed back by the inverse graph Fourier basis $\mathbf{U}^{T}$ to get the convolution result as $y=\Theta \boldsymbol{*_{\mathcal{G}}} x=\mathbf{U} \Theta \mathbf{U}^{T} x$, where $\boldsymbol{*_{\mathcal{G}}}$ is the graph convolution operator. However, all the nodes in the graph are considered  during convolution by the kernel $\Theta$ with $\mathbf{N}$ parameters. To extract spatial localization in graph, Defferrard et al. \cite{DBLP:conf/nips/DefferrardBV16} restricted the kernel $\Theta$ as a polynomial of $\mathbf{\Lambda}$  to get the $\mathbf{K}$-hop graph convolution:

\begin{footnotesize}\begin{equation}
y=\Theta(\mathbf{\Lambda}) \boldsymbol{*_{\mathcal{G}}} x= \mathbf{U}(\sum_{k=0}^{\mathbf{K}-1} \theta_{k} \mathbf{\Lambda}^{k}) \mathbf{U}^{T} x=\sum_{k=0}^{\mathbf{K}-1} \theta_{k} \mathbf{L}^{k}x
\end{equation}\end{footnotesize}where $\theta \in \mathbb{R}^{\mathbf{K}}$ is a vector of polynomial coefficients. $\mathbf{K}$ is the kernel size of graph convolution, which determines the maximum radius of the convolution from central node.

In this paper, we generalize this definition to a signal $X\in \mathbb{R}^{\mathbf{N} \times \mathbf{F}}$ with $\mathbf{F}$ input features and $\mathbf{C}$ filters as $Z=(\sum_{k=0}^{\mathbf{K}-1} \theta_{k} \mathbf{L}^{k})XW$, where $Z\in \mathbb{R}^{\mathbf{N} \times \mathbf{C}}$ is the result and $W\in \mathbb{R}^{\mathbf{F} \times \mathbf{C}}$ is a trainable parameter. The corresponding graph convolutional layer is denoted as follows:

\begin{footnotesize}\begin{equation}
\label{equ:single}
H^{(l+1)}=\boldsymbol{\rho}((\sum_{k=0}^{\mathbf{K}-1} \theta_{k} \mathbf{L}^{k}) H^{(l)} W^{(l)})\end{equation}\end{footnotesize}where $\mathbf{L}$ represents a graph based on one particular corporation relationship. $H^{(l)}\in \mathbb{R}^{\mathbf{N} \times \mathbf{F}}$ is the input at $l$-th layer and $H^{(l+1)}\in \mathbb{R}^{\mathbf{N} \times \mathbf{C}}$ is its output. $W^{(l)}\in \mathbb{R}^{\mathbf{F} \times \mathbf{C}}$  is the trainable parameter at $l^{th}$layer. $\boldsymbol{\rho}(\boldsymbol{\cdot})$ denotes the activation function, e.g. tanh, sigmoid, ReLU.

As we can see above, there are two main steps in graph convolution. The first step is to aggregate the information from surrounding stocks by multiplying Laplacian matrix and features matrix. Then a fully connected layer is implemented on the aggregated features to create high level representations for each stock. Note that with small $\mathbf{K}$, the feature aggregation will focus on close neighbors within $\mathbf{K}$ hops. Increasing the value of $\mathbf{K}$ enables model to capture larger range of cross-effect. 

\subsubsection{\textbf{Multi-Graph Convolutional Layer}}
To model the cross effect from multiple graphs, we propose the multi-graph convolution as follows:

\begin{footnotesize}\begin{equation}
\label{equ:multi}
H^{(l+1)}\!=\!\boldsymbol{\rho}((\sum_{k=0}^{\mathbf{K}-1} \theta_{k} (\theta_{S}\mathbf{L_{S}}^{k}\!+\!\theta_{I}\mathbf{L_{I}}^{k}\!+\!\theta_{T}\mathbf{L_{T}}^{k})) H^{(l)} W^{(l)})
\end{equation}\end{footnotesize}where $\{\mathbf{L_{S}},\mathbf{L_{I}},\mathbf{L_{T}}\}$ are the Laplacian matrices corresponding to adjacency matrices $\{\mathbf{A_{S}},\mathbf{A_{I}},\mathbf{A_{T}}\}$. $\{ \theta_\mathbf{{S}}, \theta_\mathbf{{I}}, \theta_\mathbf{{T}}\}$ are the trainable coefficients respectively. 

Intuitively, different relationships are likely to contribute differently for stock prediction. However, it is hard to assign them weights artificially. Therefore, we leave it to the algorithm and hope to learn the weights from data automatically. Note that our multi-graph convolution is not limited to the relationships above. It can be easily extended to incorporate more effective relationships for better stock prediction.

\subsubsection{\textbf{Dynamic Graph Convolutional Layer}}
All the pre-defined relationships require prior knowledge in financial domain and need more financial data which is unavailable sometimes. To get rid of the expert knowledge,  we design a dynamic Laplacian matrix learned by data and the corresponding layer is defined as follows:

\begin{footnotesize}\begin{equation}
\label{equ:dynamic}
H^{(l+1)}=\boldsymbol{\rho}(\mathbf{\hat{L}} H^{(l)} W^{(l)})
\end{equation}\end{footnotesize}where $\mathbf{\hat{L}}\in \mathbb{R}^{\mathbf{N} \times \mathbf{N}}$ is trainable and can be initialized just as other trainable parameters. We compare the performance of data-driven relationship with that of hand-crafted relationships in the experiments.

\subsubsection{\textbf{Multi-GCN}}
In this paper, following Kipf' work\cite{kipf2017semi}, we design our Multi-Graph Convolutional Network with two layers. The formulation of our Multi-GCN is defined as follows:

\begin{footnotesize}\begin{equation}
X^{\mathbf{GCN}}_{t}=\boldsymbol{\rho}(f(\mathbf{L})\boldsymbol{\rho}(f(\mathbf{L})X_{t}W^{1})W^{2})
\end{equation}\end{footnotesize}where $f(\mathbf{L})$ can represent one pre-defined relationship (see Equation \ref{equ:single}), or the combination of various relationships (see Equation \ref{equ:multi}), or dynamic relationship (see Equation \ref{equ:dynamic}). $X_{t}\in \mathbb{R}^{\mathbf{N} \times \mathbf{F}}$ is the input features matrix at day $t$, $X^{\mathbf{GCN}}_{t}\in \mathbb{R}^{\mathbf{N} \times \mathbf{C}}$ is the output of Multi-GCN at day $t$, which is fed into GRU later. Equal graph convolution operation with the same kernel is implemented on each day in parallel.

\subsection{Gated Reccurent Unit}
Stock prediction is a typical time-series task \cite{zhang2017stock}. RNN has shown its powerful capacity to process time-series tasks to capture long-term dependency and recent stock prediction studies have demonstrated its effectiveness \cite{chen2019investment},\cite{qin2019you}. Among various variants of RNN (e.g. vanilla RNN, Long Short Term Memory Network (LSTM), GRU), GRU is more complex than RNN and can ease gradient vanishing/exploding problems in RNN. What's more,  GRU is simpler than LSTM with fewer parameters which enables it to have shorter training time. But Chung et al. \cite{chung2014empirical} has demonstrated that GRU is as effective as LSTM empirically in many applications. Thus we choose GRU for stock price prediction.

In this paper, we consider not only the historical market data (e.g. trading prices and trading volume) but also the cross-correlation among stocks for stock movement prediction. We concatenate the high level cross-effect features produced by Multi-GCN with historical market data to form new features for prediction. These new features are put into GRU to discover temporal patterns for prediction. Our GRU hidden layer is formulated mathematically as follows:

\begin{footnotesize}\begin{equation}
\begin{split}
r_{t}&=\boldsymbol{\sigma}([H_{t-1},X_{t},X^{\mathbf{GCN}}_{t}]\boldsymbol{\cdot} W_{r}+b_{r})\\
u_{t}&=\boldsymbol{\sigma}([H_{t-1},X_{t},X^{\mathbf{GCN}}_{t}]\boldsymbol{\cdot} W_{u}+b_{u})\\
\hat{H}_{t}&=\boldsymbol{tanh}([r_{t} \boldsymbol{\odot} H_{t-1},X_{t},X^{\mathbf{GCN}}_{t}]\boldsymbol{\cdot}W_{h}+b_{h})\\
H_{t}&= u_{t} \boldsymbol{\odot} H_{t-1} + (1-u_{t}) \boldsymbol{\odot} \hat{H}_{t}\\
\end{split}
\end{equation}\end{footnotesize}where $X_{t}\in \mathbb{R}^{\mathbf{N} \times \mathbf{F}}$ is the historical records of stock collection at time $t$ and $t\in [\mathbf{d}-\mathbf{P},\cdots, \mathbf{d}]$. $X^{\mathbf{GCN}}_{t}\in \mathbb{R}^{\mathbf{N} \times \mathbf{C}}$ is the output of Multi-GCN which contains cross-effect information at time $t$. $H_{t-1}\in \mathbb{R}^{\mathbf{N} \times \mathbf{H}}$ is the hidden state at time $t-1$. $r_{t}$ is the reset gate,  $u_{t}$ is the update gate. $\boldsymbol{\sigma}\in[0,1]$ is the sigmoid activation function. Operator $\boldsymbol{\cdot}$ is the matrix multiplication, $\boldsymbol{\odot}$ is the element-wise product.
The output layer of GRU is $X_{t}^{\mathbf{GRU}}=\boldsymbol{\sigma}(H_{t}W_{g})$, where $X_{t}^{\mathbf{GRU}}\in \mathbb{R}^{\mathbf{N} \times \mathbf{G}}$, $W_{g}\in \mathbb{R}^{\mathbf{H} \times \mathbf{G}}$.

\subsection{Predictor}
The final output of GRU is $X^{\mathbf{GRU}}_{\mathbf{d}}\in \mathbb{R}^{\mathbf{N} \times \mathbf{G}}$ where $\mathbf{d}$ is the target day. A fully connected layer with sigmoid function is stacked on GRU to get the final probability prediction of stock collection. The formulation of Predictor is as follows:

\begin{footnotesize}\begin{equation}
\hat{Y}_{\mathbf{d}}=\boldsymbol{\sigma}(X^{\mathbf{GRU}}_{\mathbf{d}}W)
\end{equation}\end{footnotesize}where $\hat{Y}_{\mathbf{d}}\in \mathbb{R}^{\mathbf{N} \times \mathbf{1}}$ is the probability prediction at day $\mathbf{d}$ and $W\in \mathbb{R}^{\mathbf{G} \times \mathbf{1}}$ is trainable parameter.

\section{Experiments}
\subsection{Datasets}
To demonstrate the effectiveness of our model, we collect our datasets from a public API ((https://tushare.pro/), which are the best-known CSI (China Securities Index) 300 and CSI 500 in Chinese stock market. CSI 300 is composed of three hundred large-cap listed corporations with good liquidity. CSI 500 consists of constituent stocks chosen from top 500 mid-cap and small-cap listed companies. Their versions are defined every half a year and we fix them on 2015 January \cite{li2019multi}. Each stock in our datasets has three kinds of attributes: (1) Input Features: opening price, high price, low price, trading amount. All the input features are Min-max normalized. (2) Relationship Features: shareholder and shareholding ratio, industry category, registered capital, topicality. They are utilized to construct relational graphs. (3) Label Feature: closing price. Given the closing price of a stock at day $t$ as $P_{t}$, if $P_{t}>P_{t-1}$, we attach price movement at this day as positive with $y_{t}=1$, otherwise as negative $y_{t}=0$.

We retrieve the historical data from June 2015 to December 2019. All prices are adjusted for dividends and splits. We delete the delisted stocks during the collection period. Finally, it remains 287 stocks in CSI 300 and 489 stocks in CSI 500. To solve the problem that some stocks lack trading data for temporary suspension in some trading days, we align the historical trading days of all the stocks and fill up the missing data with trading data in most recent day. 

\begin{table}[htb]
\vspace{-1em}
\caption{The Split of Dataset}
\centering
\scriptsize
\begin{tabular}{lrrrr}
\toprule
Indexes  & Training set & Validation set & Testing set & Total \\
\midrule
CSI 500	&383,719	&54,817 	&109,633 	&548,169 \\
CSI 300	&225,209 &32,173 	&64,345 	&321,727 \\
\bottomrule
\end{tabular}
\vspace{-1em}
\end{table}
\label{tab:split}

We split the dataset into three parts: 70\% for training, then 10\% for validation and the last 20\% for testing. Details of the division of these two indexes are shown in Table above.

\subsection{Evaluation Metrics}
The stock movement prediction is a binary classification problem. Several metrics\cite{DBLP:conf/ictai/SousaSRMFM19} are selected to justify the effectiveness of all the approaches, i.e. Accuracy (ACC), Precision, Recall, F1-score and Matthews Correlation Coefficient (MCC)\cite{feng2019enhancing}. ACC measures the ratio of correct predictions over all examples. Precision focuses on the correct prediction ratio of example predicted as positive class. Recall is used to measure the fraction of positive examples that are correctly classified. F1-score is the harmonic mean of Precision and Recall. MCC can avoid bias due to data skew. All metrics are calculated on all the constituent stocks in each CSI index. The formulation of MCC is as follows:

\begin{footnotesize}\begin{equation}
\begin{split}
MCC\!\!=\!\!&\frac{TP\times TN-FP\times FN}{\sqrt{(TP\!\!+\!\!FP)\!(TP\!\!+\!\!FN)\!(TN\!\!+\!\!FP)\!(TN\!\!+\!\!FN)}}
\end{split}
\end{equation}\end{footnotesize}where TP is true positive, TN is true negative, FP is false positive, FN is false negative.

\subsection{Baselines}
The baselines in this paper can be divided into two groups. The first group only takes the historical records of the target stock as input for its movement prediction, which contains (1) Logistic Regression, (2) ARIMA (Autoregressive Integrated Moving Average)\cite{DBLP:journals/ijon/ChenH18}, (3) SVM (Support Vector Machine)\cite{DBLP:journals/asc/LuoYXP17}, (3) RF (Random Forest)\cite{DBLP:journals/eor/KraussDH17}, (4) ANN (Artificial Neural Network)\cite{DBLP:conf/ictai/LiuS17}, (5) LSTM (Long Short Term Memory)\cite{DBLP:conf/ictai/ZhaoRTS17}.

The second group considers both historical information and stock relationships, which contains (1) GCN-S: The GCN model with shareholding graph in \cite{chen2018incorporating}, (2) GCGRU-S: Our model with Shareholding Graph, (3) GCGRU-I : Our model with Industry Graph, (4) GCGRU-T: Our model with Topicality Graph, (4) Multi-GCGRU: Our model with three graphs above, (5) GCGRU-D: Our model with Dynamic Graph.

\subsection{Parameter Settings}
We set the length of lag $\mathbf{P}=5$ for all the baselines for that there are 5 trading days in a week. The optimized length of $\mathbf{P}$ is explored in our Multi-GCGRU. We train Multi-GCGRU utilizing Adam optimizer \cite{feng2019enhancing} with an initial learning rate of 0.01 and setting the mini-batch size as 32. Following Kipf' work\cite{kipf2017semi}, we set $\mathbf{K}=1$ and build our Multi-GCN with two layers. The number of corresponding hidden units are tuned within the ranges of [8, 16, 32, 64] on the validation set. The best performance is observed on [16, 32]. Our model is implemented with Tensorflow 2.1.

\subsection{Experiment Analysis}
\begin{table*}[htbp]
\caption{The Experimental results}
\centering
\scriptsize
\begin{tabular}{llrrrrr|rrrrr}
\toprule

\multirow{2}{2cm}{Input Feature}	&\multirow{2}{*}{Models}   
&\multicolumn{5}{c}{CSI300} &\multicolumn{5}{c}{CSI500}\\
&&Accuracy&Precision&Recall&F1&MCC&Accuracy&Precision&Recall&F1&MCC\\
\midrule
\multirow{6}{2cm}{Historical Records}
&LR&0.5145&0.9746&0.5133&0.6724&0.0228&0.5149&0.9723&0.5148&0.6732&0.0117\\
&SVM&0.5197&0.9498&0.5165&0.6691&0.0412&0.5253&0.9662&0.5202&0.6763&0.0636\\
&RF&0.5375&0.9298&0.5271&0.6728&0.0957&0.5433&0.9900&0.5294&0.6899&0.1587\\
&ANN&0.5191&0.9724&0.5158&0.6740&0.0463&0.5202&0.9900&0.5170&0.6792&0.0576\\
&LSTM&\textbf{0.5435}&0.9756&0.5291&0.6861&0.1443&\textbf{0.5461}&0.9662&0.5318&0.6860&0.1384\\

\midrule
		
\multirow{6}{2cm}{Historical Records \& Corporation Relationships}
&GCN-S&0.5472&0.9609&0.5317&0.6845&0.1421&0.5463&0.9675&0.5423&0.6950&0.0717\\
&GCGRU-S&0.5505&0.9321&0.5346&0.6795&0.1338&0.5521&0.9635&0.5458&0.6969&0.0938\\
&GCGRU-I&0.5598&0.9561&0.5392&0.6895&0.1739&0.5678&0.9814&0.5540&0.7082&0.1655\\
&GCGRU-T&0.5628&0.9512&0.5412&0.6899&0.1782&0.5751&0.9837&0.5581&0.7122&0.1916\\
&GCGRU-D&0.5602&0.9442&0.5402&0.6871&0.1667&0.5697&0.9844&0.5549&0.7097&0.1756\\
&Multi-GCGRU&\textbf{0.5754}&0.9603&0.5484&0.6981&0.2171&\textbf{0.5885}&0.9894&0.5658&0.7199&0.2377\\
\bottomrule
\end{tabular}
\vspace{-2em}
\label{tab:Results}
\end{table*}
In this paper, we aim to answer the following research questions:

(1) Does taking the cross effect among stocks into consideration enhance the stock movement prediction? Does our proposed model provide a better solution to incorporate the cross effect?

(2) Which kind of corporation relationship is more effective for stock prediction and why? Can we get rid of artificial relationships based on prior knowledge and learn the relationship on the basis of data? 

(3) How does our proposed Multi-GCGRU framework perform with different length of historical information?

To answer the research questions above, we conduct various experiments and deliver the final experiment results on the test datasets. Table \ref{tab:Results} answers the first and second questions and Table \ref{tab:Length} answers the third question.

\subsubsection{\textbf{Effectiveness of Cross Effect among Stocks}}
In this paper, we design two groups of approaches. The first group only considers the auto-correlation of an individual stock by taking its historical records as input. The second group considers both the auto-correlation of the target stock and cross-correlation among stocks by taking historical market data along with corporation relationships as input. By comparing the performance of these two groups, we can test the effectiveness of cross effect among stocks for prediction.

As we can see in Table \ref{tab:Results}, statistic methods perform worst for their linear and stationarity assumptions against the non-linear and dynamic properties in stock data. LSTM outperforms ANN by almost 3\% in accuracy which justifies that temporal dependency exists in stock prices. The performance of RF is nearly as better as LSTM probably due to the randomness in RF which can model the stochasticity in stock market. Besides, the GCN-S\cite{chen2018incorporating} based on shareholding relationship performs slightly better than LSTM, which indicates that the cross effect is at least as important as the temporal dependency. When considering both cross effect represented by shareholding relationship and temporal pattern in our GCGRU-S, the performance increases nearly 1\% in accuracy. All the models integrated with relational features achieve better performance than those without relationships, which proves the effectiveness of the cross effect represented by corporation relationships. 

\subsubsection{\textbf{Relationships Comparison}}
\begin{figure}[htbp]
\centering
\includegraphics[width=0.45\textwidth, height=0.15\textheight]{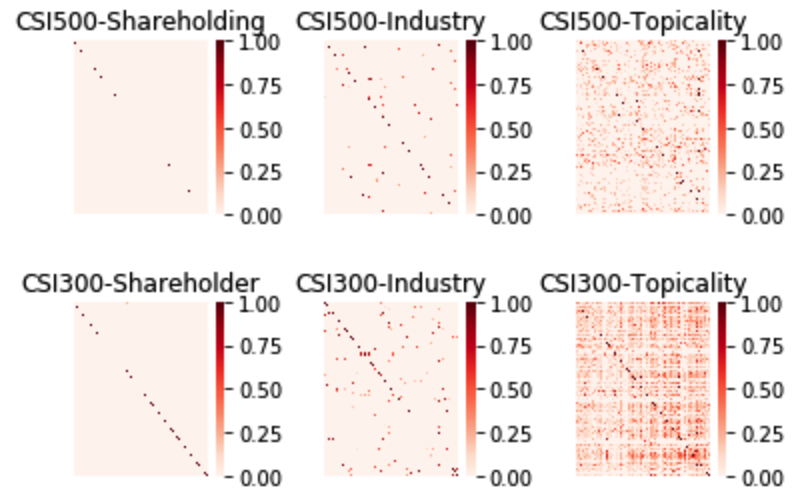}
\caption{The Visualization of Relationship Matrices}
\vspace{-1em}
\label{fig:matrix}
\end{figure}
We have pre-defined three corporation relationships based on financial knowledge, i.e. shareholding relationship, industry relationship, topicality relationship. As shown in Table \ref{tab:Results}, model fed with shareholding relationship has the worst ACC performance, while performance of model with topicality relationship is the best on both CSI300 and CSI500. The results indicate that the impact from common news on stock price is stronger than that of shareholder. This can be explained by the fact that many investors are not familiar with the shareholding structure of the corporation but they are more sensitive with public news, especially those good news and bad news. The model integrated with industry relationship also performs better than that with the shareholding relationship by at least 1\% increase of accuracy. Compared with model based on topicality relationship, it has nearly the same performance in CSI 300 and a better performance in CSI500. To further explore the correlation between relationships and their performances, we visualize the corresponding matrices (as shown in Figure \ref{fig:matrix}) and have an interesting finding that the shareholding matrix is the sparsest and the topicality is the densest. Perhaps the dense matrix contains more information helpful for prediction than the spare matrix. The result that the Multi-GCGRU with three relationships performs better than model with any single one can enhance this inference. The results above show that some relationship is more effective than other relationship, which indicates that we can further improve the performance of Multi-GCGRU through adding more effective relationship and we leave it to the interested readers.

However, all the pre-defined matrices depend on domain knowledge and extra financial data. To overcome such limitations, we explore a dynamic matrix learned by data automatically. The results show that the performance of model with data-driven matrix is between that with industry relation and topicality relation. Although the model with data-driven relationship does not have the best performance, it can also be a choice when financial knowledge and data are insufficient.

\subsubsection{\textbf{The Length of Historical Information}}
\begin{table}[htbp]
\vspace{-1em}
\caption{Multi-GCGRU with Different Lag Sizes}
\centering
\footnotesize
\begin{tabular}{lrr|rrr}
\toprule
\multirow{2}{*}{Length}	&\multicolumn{2}{c}{CSI300}	&\multicolumn{2}{c}{CSI500}\\
&ACC&MCC&ACC&MCC\\
\midrule
3-days	&0.5623			&0.1513	&0.5752			&0.1742\\
5-days	&0.5754			&0.2171	&0.5885			&0.2377\\
7-days	&\textbf{0.5790}	&0.2196	&\textbf{0.5901}	&0.2821\\
9-days	&0.5769			&0.1869	&0.5783			&0.1965\\
11-days	&0.5705			&0.1378	&0.5691			&0.1221\\					
\bottomrule
\end{tabular}
\vspace{-2em}
\label{tab:Length}
\end{table}
\begin{figure}[htbp]
\centering
\includegraphics[width=0.4\textwidth, height=0.2\textheight]{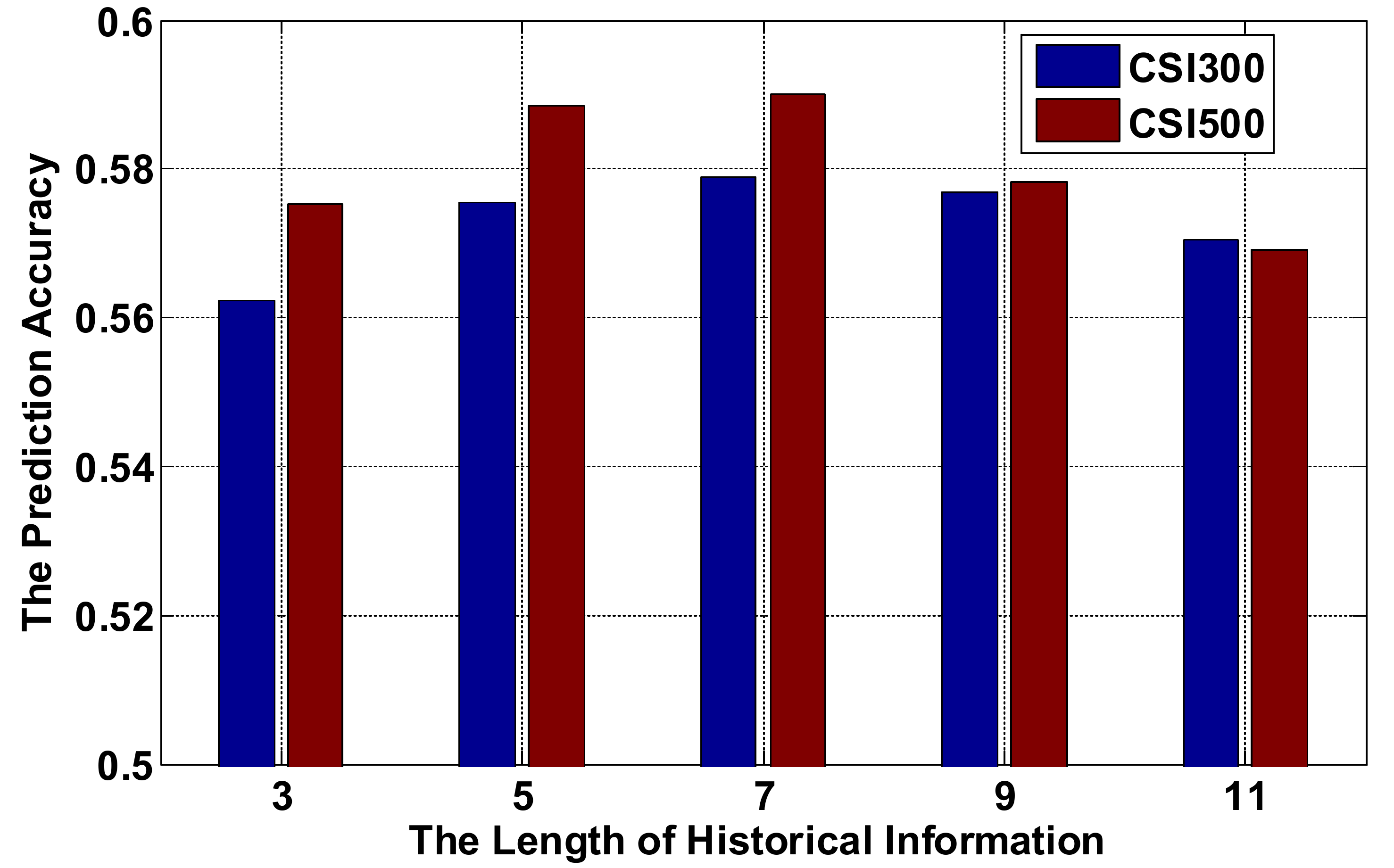}
\caption{Accuracy of Multi-GCGRU with Different Lag Sizes}
\vspace{-0.5em}
\label{fig:Bar}
\end{figure}

We conduct experiments on different length of days, specifically [3, 5, 7, 9, 11] previous days. As shown in Table \ref{tab:Length} and Figure \ref{fig:Bar}, our model achieves the best performance at 7 days , with the accuracy 57.90\% on CSI300 dataset and 59.01\% on CSI500. The worst performance occurs at 3-days with accuracy 56.23\% on CSI300 and  at 11-days with accuracy 56.91\% on CSI500. Therefore, the length of historical information has an impact on prediction performance.
\section{Conclusion}
The price movement of an individual stock is inevitably influenced by other related stocks. This paper justifies that taking cross effect among stocks can effectively improve the prediction accuracy. Our contribution is that we model the cross effect by encoding various corporation relationships into graphs, based on which we propose a Multi-GCGRU framework to capture both auto-correlation and cross-correlation properties in stock prices. We first employ Multi-GCN to extract cross effect among stocks based on three pre-defined graphs. Then we utilize GRU to process the cross-effect features produced by Multi-GCN along with historical market information to model temporal pattern in stock prices. Further, we explore a data-driven graph to overcome dependency on prior financial knowledge. Our model can be easily extended to incorporate more effective relationships among stocks. 
\section{Acknowledgment}
The authors would like to thank reviewers for their valuable comments.This work is supported by the National Key R\&D Program of China (No. 2019YFB2102100), National Natural Science Foundation of China (No. 61802387, No.61806192), Science and Technology Development Fund of Macao S.A.R (FDCT) under number 0015/2019/AKP, Shenzhen Discipline Construction Project for Urban Computing and Data Intelligence, China's Post-doctoral Science Fund (No.2019M663183), and National Natural Science Foundation of Shenzhen (No.JCYJ20190812153212464, No.JCYJ20190812160003719).

\vspace{-0.2cm}

\bibliographystyle{IEEEtran}
\bibliography{main}

\begin{thebibliography}{10}
\providecommand{\url}[1]{#1}
\csname url@samestyle\endcsname
\providecommand{\newblock}{\relax}
\providecommand{\bibinfo}[2]{#2}
\providecommand{\BIBentrySTDinterwordspacing}{\spaceskip=0pt\relax}
\providecommand{\BIBentryALTinterwordstretchfactor}{4}
\providecommand{\BIBentryALTinterwordspacing}{\spaceskip=\fontdimen2\font plus
\BIBentryALTinterwordstretchfactor\fontdimen3\font minus
  \fontdimen4\font\relax}
\providecommand{\BIBforeignlanguage}[2]{{%
\expandafter\ifx\csname l@#1\endcsname\relax
\typeout{** WARNING: IEEEtran.bst: No hyphenation pattern has been}%
\typeout{** loaded for the language `#1'. Using the pattern for}%
\typeout{** the default language instead.}%
\else
\language=\csname l@#1\endcsname
\fi
#2}}
\providecommand{\BIBdecl}{\relax}
\BIBdecl

\bibitem{HouIndustry}
Hou and Kewei, ``Industry information diffusion and the lead-lag effect in
  stock returns,'' \emph{The Review of Financial Studies}, vol.~20, no.~4, pp.
  1113--1138, 2007.

\bibitem{DBLP:journals/asc/LuoYXP17}
L.~Luo, S.~You, Y.~Xu, and H.~Peng, ``Improving the integration of piece wise
  linear representation and weighted support vector machine for stock trading
  signal prediction,'' \emph{Appl. Soft Comput.}, vol.~56, pp. 199--216, 2017.

\bibitem{DBLP:journals/asc/ChangL17}
Y.~Chang and M.~Lee, ``Incorporating markov decision process on genetic
  algorithms to formulate trading strategies for stock markets,'' \emph{Appl.
  Soft Comput.}, vol.~52, pp. 1143--1153, 2017.

\bibitem{jin2017tracking}
F.~Jin, W.~Wang, P.~Chakraborty, N.~Self, F.~Chen, and N.~Ramakrishnan,
  ``Tracking multiple social media for stock market event prediction,'' in
  \emph{ICDM}, 2017, pp. 16--30.

\bibitem{DBLP:conf/ictai/ChangT18}
J.~Chang and W.~Tu, ``A stock-movement aware approach for discovering
  investors' personalized preferences in stock markets,'' in \emph{{ICTAI}},
  2018, pp. 275--280.

\bibitem{feng2019enhancing}
F.~Feng, H.~Chen, X.~He, J.~Ding, M.~Sun, and T.-S. Chua, ``Enhancing stock
  movement prediction with adversarial training,'' in \emph{Proceedings of the
  28th International Joint Conference on Artificial Intelligence}, 2019, pp.
  5843--5849.

\bibitem{DBLP:conf/ictai/LiuS17}
H.~Liu and B.~Song, ``Stock trends forecasting by multi-layer stochastic {ANN}
  bagging,'' in \emph{{ICTAI}}, 2017, pp. 322--329.

\bibitem{zhang2017stock}
L.~Zhang, C.~Aggarwal, and G.-J. Qi, ``Stock price prediction via discovering
  multi-frequency trading patterns,'' in \emph{SIGKDD}, 2017, pp. 2141--2149.

\bibitem{chen2019investment}
C.~Chen, L.~Zhao, J.~Bian, C.~Xing, and T.-Y. Liu, ``Investment behaviors can
  tell what inside: Exploring stock intrinsic properties for stock trend
  prediction,'' in \emph{SIGKDD}, 2019, pp. 2376--2384.

\bibitem{Andrew1990When}
A.~W. Lo and A.~C. MacKinlay, ``When are contrarian profits due to stock market
  overreaction?'' \emph{The Review of Financial Studies}, 1990.

\bibitem{DBLP:journals/epjds/LetiziaL19}
E.~Letizia and F.~Lillo, ``Corporate payments networks and credit risk
  rating,'' \emph{{EPJ} Data Sci.}, vol.~8, no.~1, pp. 21:1--21:29, 2019.

\bibitem{chen2018incorporating}
Y.~Chen, Z.~Wei, and X.~Huang, ``Incorporating corporation relationship via
  graph convolutional neural networks for stock price prediction,'' in
  \emph{CIKM}, 2018, pp. 1655--1658.

\bibitem{DBLP:conf/wsdm/HuLBLL18}
Z.~Hu, W.~Liu, J.~Bian, X.~Liu, and T.~Liu, ``Listening to chaotic whispers: A
  deep learning framework for news-oriented stock trend prediction,'' in
  \emph{{WSDM}}, 2018, pp. 261--269.

\bibitem{duvenaud2015convolutional}
D.~K. Duvenaud, D.~Maclaurin, J.~Iparraguirre, R.~Bombarell, T.~Hirzel,
  A.~Aspuru-Guzik, and R.~P. Adams, ``Convolutional networks on graphs for
  learning molecular fingerprints,'' in \emph{Advances in neural information
  processing systems}, 2015, pp. 2224--2232.

\bibitem{ying2018graph}
R.~Ying, R.~He, K.~Chen, P.~Eksombatchai, W.~L. Hamilton, and J.~Leskovec,
  ``Graph convolutional neural networks for web-scale recommender systems,'' in
  \emph{SIGKDD}, 2018, pp. 974--983.

\bibitem{cuitraffic19}
Z.~Cui, K.~Henrickson, R.~Ke, and Y.~Wang, ``Traffic graph convolutional
  recurrent neural network: A deep learning framework for network-scale traffic
  learning and forecasting,'' \emph{IEEE Transactions on Intelligent
  Transportation Systems}, vol.~PP, 2019.

\bibitem{chung2014empirical}
J.~Chung, C.~Gulcehre, K.~Cho, and Y.~Bengio, ``Empirical evaluation of gated
  recurrent neural networks on sequence modeling,'' in \emph{NIPS Workshop},
  2014.

\bibitem{DBLP:conf/sdm/KampBG14}
M.~Kamp, M.~Boley, and T.~Gartner, ``Beating human analysts in nowcasting
  corporate earnings by using publicly available stock price and correlation
  features,'' in \emph{ICDM}, 2014, pp. 641--649.

\bibitem{CHAN2003223}
W.~S. Chan, ``Stock price reaction to news and no-news: drift and reversal
  after headlines,'' \emph{Journal of Financial Economics}, vol.~70, no.~2, pp.
  223 -- 260, 2003.

\bibitem{DBLP:conf/acl/RekabsazLBDAH17}
N.~Rekabsaz, M.~Lupu, A.~Baklanov, A.~Dur, L.~Andersson, and A.~Hanbury,
  ``Volatility prediction using financial disclosures sentiments with word
  embedding-based {IR} models,'' in \emph{{ACL}}, 2017, pp. 1712--1721.

\bibitem{DBLP:conf/ictai/SousaSRMFM19}
M.~G. Sousa, K.~M. Sakiyama, L.~de~Souza~Rodrigues, P.~H. Moraes, E.~R.
  Fernandes, and E.~T. Matsubara, ``{BERT} for stock market sentiment
  analysis,'' in \emph{{ICTAI}}, 2019, pp. 1597--1601.

\bibitem{li2015tensor}
Q.~Li, L.~Jiang, P.~Li, and H.~Chen, ``Tensor-based learning for predicting
  stock movements,'' in \emph{AAAI}, 2015.

\bibitem{xu2018stock}
Y.~Xu and S.~B. Cohen, ``Stock movement prediction from tweets and historical
  prices,'' in \emph{ACL}, 2018, pp. 1970--1979.

\bibitem{bordino2014stock}
I.~Bordino, N.~Kourtellis, N.~Laptev, and Y.~Billawala, ``Stock trade volume
  prediction with yahoo finance user browsing behavior,'' in \emph{ICDE}, 2014,
  pp. 1168--1173.

\bibitem{ding2015deep}
X.~Ding, Y.~Zhang, T.~Liu, and J.~Duan, ``Deep learning for event-driven stock
  prediction,'' in \emph{Twenty-fourth international joint conference on
  artificial intelligence}, 2015.

\bibitem{qin2019you}
Y.~Qin and Y.~Yang, ``What you say and how you say it matters: Predicting
  financial risk using verbal and vocal cues,'' in \emph{ACL}, 2019, p. 390.

\bibitem{li2019multi}
C.~Li, D.~Song, and D.~Tao, ``Multi-task recurrent neural networks and
  higher-order markov random fields for stock price movement prediction:
  Multi-task rnn and higer-order mrfs for stock price classification,'' in
  \emph{SIGKDD}, 2019, pp. 1141--1151.

\bibitem{kipf2017semi}
T.~N. Kipf and M.~Welling, ``Semi-supervised classification with graph
  convolutional networks,'' in \emph{{ICLR}}, 2017, pp. 1--12.

\bibitem{9207049}
J.~{Ye}, J.~{Zhao}, K.~{Ye}, and C.~{Xu}, ``Multi-stgcnet: A graph convolution
  based spatial-temporal framework for subway passenger flow forecasting,'' in
  \emph{2020 International Joint Conference on Neural Networks (IJCNN)}, 2020,
  pp. 1--8.

\bibitem{DBLP:journals/corr/abs-2005-11691}
J.~Ye, J.~Zhao, K.~Ye, and C.~Xu, ``How to build a graph-based deep learning
  architecture in traffic domain: {A} survey,'' \emph{CoRR}, vol.
  abs/2005.11691, 2020.

\bibitem{ATTIG20062875}
N.~Attig, W.-M. Fong, Y.~Gadhoum, and L.~H. Lang, ``Effects of large
  shareholding on information asymmetry and stock liquidity,'' \emph{Journal of
  Banking \& Finance}, vol.~30, no.~10, pp. 2875 -- 2892, 2006.

\bibitem{Hoffmann2013426}
M.~Hoffmann, M.~Rosenbaum, and N.~Yoshida, ``Estimation of the lead-lag
  parameter from non-synchronous data,'' \emph{Bernoulli}, vol.~19, no.~2, pp.
  426--461, 2013.

\bibitem{DBLP:conf/nips/DefferrardBV16}
M.~Defferrard, X.~Bresson, and P.~Vandergheynst, ``Convolutional neural
  networks on graphs with fast localized spectral filtering,'' in \emph{NIPS},
  2016, pp. 3837--3845.

\bibitem{DBLP:journals/corr/BrunaZSL13}
J.~Bruna, W.~Zaremba, A.~Szlam, and Y.~LeCun, ``Spectral networks and locally
  connected networks on graphs,'' in \emph{{ICLR}}, 2014.

\bibitem{DBLP:journals/ijon/ChenH18}
Y.~Chen and Y.~Hao, ``Integrating principle component analysis and weighted
  support vector machine for stock trading signals prediction,''
  \emph{Neurocomputing}, vol. 321, pp. 381--402, 2018.

\bibitem{DBLP:journals/eor/KraussDH17}
C.~Krauss, X.~A. Do, and N.~Huck, ``Deep neural networks, gradient-boosted
  trees, random forests: Statistical arbitrage on the s{\&}p 500,'' \emph{Eur.
  J. Oper. Res.}, vol. 259, no.~2, pp. 689--702, 2017.

\bibitem{DBLP:conf/ictai/ZhaoRTS17}
Z.~Zhao, R.~Rao, S.~Tu, and J.~Shi, ``Time-weighted {LSTM} model with redefined
  labeling for stock trend prediction,'' in \emph{{ICTAI}}, 2017, pp.
  1210--1217.

\end{thebibliography}
\end{document}